\documentclass[final,3p,11pt]{elsarticle}

\usepackage{latexsym}
\usepackage{graphicx}
\usepackage{bbm}
\usepackage{amsmath,amssymb, mathrsfs}
\usepackage[latin1]{inputenc}
\usepackage[dvips]{hyperref}

\newcommand{\qs}{Q_\mathrm{s}}

\newcommand{\as}{{\alpha_\mathrm{s}}}
\newcommand{\rt}{\boldsymbol{r}_\perp}
\newcommand{\xt}{\boldsymbol{x}_\perp}
\newcommand{\pt}{{\boldsymbol{p}_\perp}}

\newcommand{\kt}{\boldsymbol{k}_\perp}
\newcommand{\nc}{{N_\mathrm{c}}}

\newcommand{\ca}{{C_\mathrm{A}}}

\newcommand{\npart}{{N_\textrm{part}}}

\newcommand{\nft}{N_{_{\rm FT}}}

\newcommand{\nr}[1]{(\ref{#1})} 
\newcommand{\ud}{\mathrm{d}}
\newcommand{\fig}{fig.~}

\newcommand{\eq}{eq.~}

\def\p{{\boldsymbol p}}

\begin{document}

\begin{frontmatter}

\title{Glittering Glasmas}
\author[cea]{F. Gelis}
\author[cea,jyu]{T. Lappi}
\author[cea,bnl]{L. McLerran}
\address[cea]{Institut de Physique Th\'eorique, %
B\^at. 774, CEA/DSM/Saclay, 91191 Gif-sur-Yvette, France}
\address[jyu]{Department of Physics, %
 P.O. Box 35, 40014 University of Jyv\"askyl\"a, Finland}
\address[bnl]{Physics Department and Riken Brookhaven Center, Brookhaven
National Laboratory, Upton, NY 11973, USA}

\begin{abstract}
  We compute the production of gluons from Glasma color flux tubes.
  We calculate the probability distribution of gluon multiplicities
  arising from the distribution of color electric and color magnetic
  flux tubes found in the Glasma.  We show that the result corresponds
  to the negative binomial probability distribution observed in
  experiments. The parameter $k$ that characterizes this distribution
  is proportional to the number of colors $\nc^2-1$ and to the number
  of flux tubes. For one gluon color and one flux tube, the
  multiplicity distribution is close to a Bose-Einstein distribution.
  We call this decay process ``Glitter'', a term that is explained
  below.
\end{abstract}


\end{frontmatter}

\section{Introduction}
In high energy nuclear collisions, it has been argued that a Glasma is
formed through the collision of two sheets of Color Glass Condensate
(CGC)~\cite{McLerran:1994ni,McLerran:1994ka,McLerran:1994vd,Kovchegov:1996ty,%
  Jalilian-Marian:1997xn,Jalilian-Marian:1997jx,Jalilian-Marian:1997gr,%
  Iancu:2000hn,%
  Ferreiro:2001qy,Iancu:2001md,Mueller:2001uk,Kovner:1995ts,Kovner:1995ja,%
  Krasnitz:1998ns,Krasnitz:2000gz,Krasnitz:2001qu,Lappi:2003bi,Lappi:2006fp} (for a
review and additional references see
\cite{Iancu:2003xm,Weigert:2005us}).  At very high energies, the gluon
density per unit area per unit rapidity, $\ud N/\ud y \ud^2\rt$, is of
order $\qs^2/\as$ in the CGC and Glasma.  This density is so large
that the interaction strength of QCD is weak, $\as \ll 1$.  Indeed, in
these relations, the scale of the coupling constant is set by the
saturation momentum $\qs$, and the saturation momentum grows with both
increasing energy and size of the nucleus.  The Glasma is formed
during the time it takes two Lorentz contracted sheets of Colored
Glass to pass through one another, $\tau \sim e^{-\kappa/\as}/\qs$ a
time parametrically short compared to the natural time scale for decay
of the flux tubes, $\tau_\textrm{decay} \sim 1/\qs$.  We shall not
describe in detail the properties of either the Glasma or the CGC in
this paper and refer the reader to the original literature for
details.

In this paper, we compute the probability distribution for the
multiplicity of gluons produced in the Glasma.  We show that the
distribution of gluons arising from such a decay is not a Poisson
distribution as might be expected for the decay of an external source
into particles in a weakly interacting theory.  It turns out the
distribution is a negative binomial distribution.

Recall that a Poisson distribution,
\begin{equation}
P^\textrm{Poisson}_n = {1 \over {n!}} \bar{n}^n e^{-\bar{n}} 
\end{equation}
is completely characterized by its mean value $\bar{n}$. 
In contrast, a negative binomial distribution is a 2-parameter
distribution of the form
\begin{equation}
P_n^{^{\rm NB}}= 
\frac{\Gamma(k+n)}{\Gamma(k)\Gamma(n+1)} 
\frac{\bar{n}^n k^k}{(\bar{n}+k)^{n+k}}
\; .
\end{equation}
This distribution has larger fluctuations than a Poisson
distribution, but tends to a Poisson distribution if $k\to +\infty$ at
fixed $\bar{n}$.

For the processes we consider, the decays will be into one coherent
state associated with a gluon. 
For this reason, and because a negative binomial distribution does not fall
as $1/n!$ at large $n$, as does a Poisson distribution, we will refer to 
this property of the distribution as
tenacious.  The tenaciousness of this distribution results in an
amplification of the intensity of multiply emitted gluons relative to
that of a Poisson distribution.  This means that there are larger
fluctuations in high gluon multiplicity events than would be typical
of a Poisson distribution.  Thus, we shall use the acronym ``Glitter''
to describe radiation from these flux tubes, as an abbreviation for
GLuon Intensification Through Tenacious Emission of Radiation.

In this paper, we shall compute the multiplicity distribution of
gluons produced by Glasma flux tubes.  We show that a single flux tube
decays into gluons with a negative binomial distribution and is
characterized by a parameter $k_0$ of order one.  This implies that a
decay of $\nft$ flux tubes produces a negative binomial distribution
characterized by $k = \nft k_0$.  For $k = 1$, a negative
binomial distribution is a Bose-Einstein distribution\footnote{A
  Bose-Einstein, or geometrical, distribution is a thermal distribution for single
  state systems. Its probability distribution reads:
\begin{equation*}
P_n^{^{\rm BE}}=\frac{1}{1+\bar{n}}\left(\frac{\bar{n}}{1+\bar{n}}\right)^n\; .
\end{equation*}},
so single flux tube decays are close in form to Bose-Einstein distributions.
In addition to the parameter $k$, the negative binomial distribution
is also parameterized by the average multiplicity $\bar{n}$. The ratio
$\bar{n}/k = \bar{n}/\nft k_0$ is approximately the multiplicity per
flux tube.
 
Based on this interpretation of the decay of an ensemble of flux tubes
we argue that the flux tubes are a ``glittering'' glasma.  We will
then  iscuss our results in the context of the extraction of negative
binomial distribution from the UA(5) and PHENIX collaborations.

\section{Calculation of the multiplicity distribution}
The probability distribution of a discrete quantity is conveniently
defined in terms of its generating function
\begin{equation}
F(z) \equiv \sum_{n=0}^{\infty} z^n P_n,
\end{equation}
from which one can compute the moments of the multiplicity distribution as
\begin{equation}\label{eq:defmoment}
\langle n(n-1)\cdots (n-q+1) \rangle = 
\left. \frac{\ud^q F(z)}{\ud z^q}\right|_{z=1} .
\end{equation}
(The moments defined in this way are known as the factorial moments.)
It was shown in Refs.~\cite{Gelis:2008rw,Gelis:2008ad,Gelis:2008sz}
that in the case of the central rapidity region of nucleus-nucleus
collisions, when both nuclei can be described as strong color sources
$\rho \sim 1/g$, these moments can be computed as
\begin{equation}\label{eq:moment}
\langle n(n-1)\cdots (n-q+1) \rangle = 
\int [\ud \rho_1][\ud \rho_2] W_y[\rho_1] W_y[\rho_2] \big(n[\rho_1,\rho_2]\big)^q.
\end{equation}
This result is valid to leading log accuracy, i.e. it includes the
leading order in $\as$ with all the powers of $\as \ln 1/x$ resummed
into the rapidity dependence of the weight functionals $W_y[\rho_1]$.
The factor $n[\rho_1,\rho_2]$ inside the integral in the r.h.s.
of eq.~\nr{eq:moment} is the integrated multiplicity corresponding to a
fixed configuration of color charge densities $\rho_1$ and $\rho_2$:
\begin{equation}\label{eq:single}
n[\rho_1,\rho_2] = \int \ud^2 \pt \ud y_p \frac{\ud N}{\ud^2 \pt \ud y_p}\; ,
\end{equation}
where ${\ud N}/{\ud^2 \pt \ud y_p}$ is obtained by Fourier
transforming the classical gauge field radiated by the sources
$\rho_{1,2}$. In our leading log calculation the classical fields and
thus also the single gluon multiplicity are boost invariant, so the
integral over rapidity in \eq\nr{eq:single} is just a constant factor.
We are assuming that the rapidity interval is small enough compared to
$1/\as$; otherwise there are additional large logarithms that must be
resummed; this case is studied in detail in Ref.~\cite{Gelis:2008sz}.

In the Glasma the single inclusive multiplicity is of order $1/\as$
and thus the moment defined in \eq\nr{eq:moment} is of order
$(1/\as)^{q}$. We are computing the moments only to leading order in
$\as$, thus powers of $n$ lower than $q$ are negligible compared to
the $q$'th moment. In particular $\langle n(n-1)\cdots (n-q+1) \rangle
\approx \langle n^q \rangle$ at this level of accuracy.  This means
that the leading (in $\as$) correlations in the multiplicity
distribution come entirely from the average over the distribution of
sources $\rho_{1,2}$; i.e.  the dominant behavior of the probability
distribution comes from the large logarithms of the energy resummed
into the $W$'s.  The contributions to the probability distribution for
fixed sources from higher loop orders that were studied in
Ref.~\cite{Gelis:2006yv} are suppressed by powers of $\as$ and thus
contribute to our calculation only when they are enhanced by large
logarithms.

\begin{figure}
\begin{center}
\includegraphics[width=4cm]{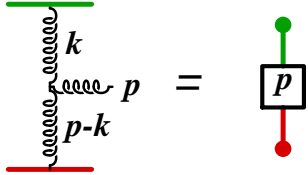}
\end{center}
\caption{The fundamental building block, the one gluon production amplitude.}
\label{fig:block}
\end{figure}

\begin{figure}
\begin{center}
\includegraphics[width=6cm]{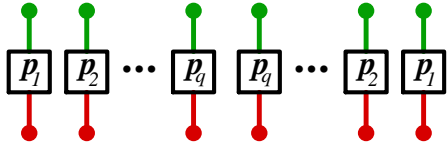}
\end{center}
\caption{Diagrams that have to be contracted into connected one.}
\label{fig:contractionproblem}
\end{figure}

The fundamental properties of the probability distribution are better
reflected in the factorial cumulants\footnote{Note that the
  \emph{factorial} cumulant defined here differs from the conventional
  one in that we are defining the cumulant from $\langle n(n-1)\cdots
  (n-q+1) \rangle$ instead of $\langle n^q\rangle$. In terms of the
  generating function this is a question of differentiating w.r.t. $z$
  in stead of $\ln z$. As explained above the difference between the
  two is of higher order in $\as$ than our calculation and we will not
  discuss it further.}
\begin{equation}\label{eq:defcumul}
m_q \equiv \langle n(n-1)\cdots (n-q+1) \rangle - \textrm{ disc. } = 
\left. \frac{\ud^q \ln F(z)}{\ud z^q}\right|_{z=1}\;,
\end{equation}
where ``disc.'' denotes the disconnected contributions that can be
expressed in terms of the lower cumulants.  We shall now turn to
calculating these quantities of the multiplicity distribution in the
Glasma at the lowest nontrivial order in the sources. This
approximation is equivalent to assuming that the momenta of the
produced gluons are all larger than the saturation scale. For
concreteness one can take the distribution of the sources from the MV
model:
\begin{equation}\label{eq:MV}
W[\rho] = \mathcal{C} \,\exp \left[- \int \ud^2 \xt \frac{\rho^a(\xt)\rho^a(\xt)}{g^4 \mu^2} \right]\; .
\end{equation}
Since the probability distribution essentially depends only on the
combinatorics of pairwise source connections, our result applies
equally well to a nonlocal Gaussian distribution that would more
closely reproduce a solution of the BK 
equation~\cite{Balitsky:1995ub,Kovchegov:1996ty,Kovchegov:1999ua}.

The calculation proceeds in the same way as that of the second and
third cumulants computed in
Refs.\cite{Dumitru:2008wn,Gavin:2008ev,Dusling:2009ar}, and we refer
to these works for a more detailed description.  Computing the
probability distribution to all orders in the sources $\rho_{1,2}$ is
in principle possible using the methods developed for the single
inclusive gluon
production~\cite{Krasnitz:1998ns,Krasnitz:2000gz,Krasnitz:2001qu,Lappi:2003bi},
but reliably computing the higher cumulants requires a significant
numerical effort to gather enough statistics and is left for future
work.

\begin{figure}[t]
\begin{center}
\includegraphics[width=6cm]{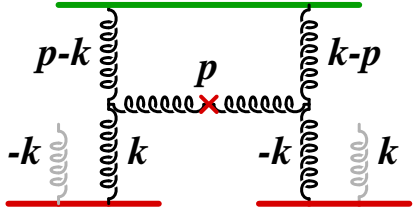}
\end{center}
\caption{Contraction contributing to the dominant correlation, building 
block of rainbow diagram 
}
\label{fig:rainbow}
\end{figure}

\begin{figure}[t]
\begin{center}
\includegraphics[width=6cm]{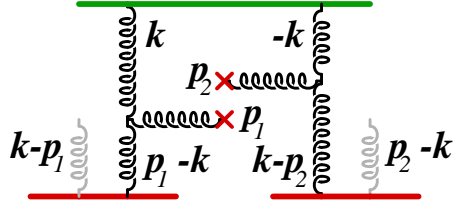}
\end{center}
\caption{Contraction contributing to a subdominant correlation, non-rainbow diagram}
\label{fig:nonrainbow}
\end{figure}

The fundamental building block in our calculation is the amplitude to
produce one gluon with momentum $\p$ from the fixed classical color
charges $\rho_1(\kt)$ and $\rho_2(\pt-\kt)$. This amplitude reads
\begin{equation}
\frac{\rho_1(\kt)}{\kt^2}
\frac{\rho_2(\pt-\kt)}{(\pt-\kt)^2}\;
L^\gamma(\p,\kt)\; ,
\end{equation}
where we are not writing the color indices explicitly.  Here,
$L^\gamma(\p,\kt)$ denotes the effective Lipatov vertex. We do not
need the explicit expression of its components, and it will be
sufficient to know that it satisfies the following two properties
\begin{eqnarray}
L^\gamma(\p,\kt) &=& L^\gamma(\p,\pt - \kt)\; ,
\nonumber\\
L^\gamma(\p,\kt)L_\gamma(\p,\kt) &=& - 4 \frac{(\pt-\kt)^2\kt^2}{\pt^2}\; .
\end{eqnarray}
The diagrammatic notation for this amplitude is shown in
\fig\ref{fig:block}.  To compute the $q$'th cumulant we need to take
$2q$ factors of this basic building block ($q$ for the amplitude and
$q$ for the complex conjugate) and perform the averages over the
sources.  Because the distribution of the sources in \eq\nr{eq:MV} is
Gaussian, we only need to keep track of contractions of pairs of
sources $\rho_1$ and (separately) of pairs of sources $\rho_2$ and
replace them by the correlator
\begin{equation}\label{eq:rhorho}
\langle \rho(\kt)  \rho(\kt')\rangle = (2 \pi)^2 \delta^2(\kt + \kt') g^4\mu^2(\kt).
\end{equation}
For the time being we shall leave an unspecified $\kt$-dependence in 
the correlation function $g^4\mu^2(\kt)$.
In the MV model~\cite{McLerran:1994ni,McLerran:1994ka,McLerran:1994vd}
$g^4\mu^2(\kt)$ is a constant, but
JIMWLK or BK evolution can effectively lead to a different 
$\kt$-dependence~\cite{Iancu:2002aq}. 
With the simplified diagrammatic notation introduced in
\fig\ref{fig:block} this combinatoric problem now corresponds to
forming a connected contraction of the $2q$ boxes, each with two lines
attached, illustrated in \fig\ref{fig:contractionproblem}.

\begin{figure}[t]
\begin{center}
\includegraphics[width=5cm]{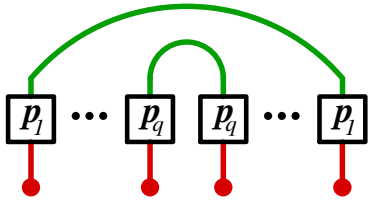}
\end{center}
\caption{Rainbow diagram.}
\label{fig:simplerainbow}
\end{figure}

\begin{figure}[t]
\begin{center}
\includegraphics[width=5cm]{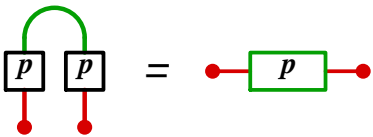}
\end{center}
\caption{Dimer notation for rainbow-like links.}
\label{fig:dimer}
\end{figure}

We shall now show why the only contributing contractions are
``rainbow'' diagrams, where on one side (upper or lower) of the
diagram the two boxes corresponding to the same momentum $\p_r$ are
contracted with each other, as in \fig\ref{fig:simplerainbow}. An
example of a building block of a rainbow diagram is given in
\fig\ref{fig:rainbow}. There are a total of four propagators with the
same momentum $\kt$. Two of these are cancelled by the contractions of
the four Lipatov vertices attached to the ends, leaving a
quadratically infrared divergent contribution $\sim \ud^2 \kt/\kt^4$
to the integral over $\kt$.  These kinds of divergences are a sign
that this contribution to the multiparticle correlation is sensitive
to the whole correlated area in the transverse plane.  They are
regulated at the scale $\qs$ (since $\qs^{-1}$ is the correlation
length between color charges in the transverse direction), giving a
contribution of order $1/\qs^2$.  Compare this to the ``non-rainbow''
contribution depicted in \fig\ref{fig:nonrainbow}. Here there are only
two propagators with the same momenta and the $\kt$-integral is
convergent. Instead of ${\cal O}(1/\qs^2)$, this yields a contribution
$\sim 1/\pt_1^2$ or $\sim 1/\pt_2^2$ which shall be neglected here
since we are assuming $\pt \gg \qs$. Thus the contractions of the
boxes in \fig\ref{fig:contractionproblem} have to form a ``rainbow
diagram'' (\fig\ref{fig:simplerainbow}) in either the upper or lower
part of the diagram. There cannot be a rainbow on both sides since
this would lead to a disconnected contribution.

\begin{figure}[t]
\begin{center}
\includegraphics[width=0.6\textwidth]{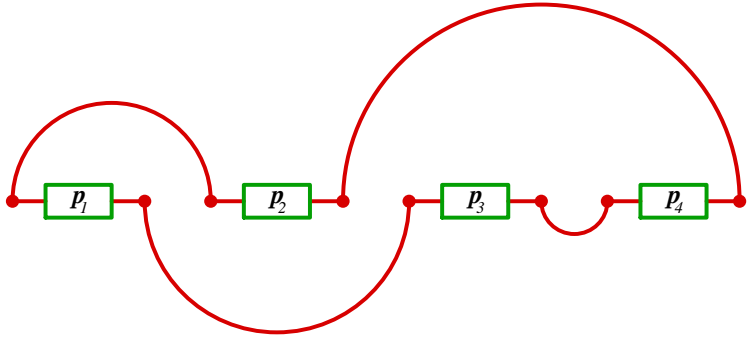}
\end{center}
\caption{Connected diagram in the polymer notation introduced in 
\fig\ref{fig:dimer}}
\label{fig:polymer}
\end{figure}

Now that we have reduced the combinatoric problem to rainbow diagrams
we can simplify our diagrammatic notation even further, as shown in
\fig\ref{fig:dimer}. We can consider the two boxes connected by the
line in the rainbow as a dimer with two ends. Now we must count the
number of ways in which our $q$ dimers can be combined to form a
connected loop. The first end of the first dimer can be connected to
$2q-2$ other loose dimer ends, let us assume it is connected to dimer
$r$. The second end of dimer $r$ now has $2q-4$ loose ends available,
because closing the loop with the first dimer would immediately give a
disconnected contribution. Thus we arrive at
$(2q-2)\cdot(2q-4)\cdot\cdots\cdot2 = 2^{q-1}(q-1)!$ combinations.
Remembering that we can form the rainbow on either the upper or lower
side of the diagram we finally arrive at $2^q(q-1)!$ topologies. In
the special cases of $q=2$ and $q=3$ this gives 4 and 16, which were
the number of diagrams evaluated in \cite{Dumitru:2008wn} and
\cite{Dusling:2009ar}.

The color structure can be simplified in the following way. Let us
denote the color indices of the sources $i=1\dots q$ contracted on the
upper side of the diagram in \fig\ref{fig:simplerainbow} by $a_i$
(because of the rainbow structure of the connections on the upper
side, we do not need separate indices for the sources in the complex
conjugate amplitude), the color index of the gluon with $\p_i$ by
$b_i$ and the color indices on the lower side of the diagram $c_i$
(source connected to gluon $\p_i$ in the amplitude) and $c'_i$ (in the
complex conjugate). Now the color structure of the upper side of the
rainbow and the boxes is $f^{a_1 b_1 c_1}f^{a_1 b_1 c'_1} \dots f^{a_q
  b_q c_q}f^{a_q b_q c'_q} = (\ca)^q \delta^{c_1 c'_1}\dots
\delta^{c_q c'_q}$. $\ca=\nc$ is the Casimir of the adjoint
representation. The $c_i,c_i^\prime$ indices must now be contracted
pairwise into a single connected loop as in \fig\ref{fig:polymer},
which yields a factor ${\rm tr}(1_{\rm adj})=\nc^2-1$, making the
total color factor $(\nc)^q(\nc^2-1)$.

Let us then turn to the structure of momentum flow in the diagram.
Transverse momentum is conserved at the vertices and in the sources
connections due to the expectation value \eq\nr{eq:rhorho}. Altogether
there are originally $4q$ transverse momentum integrals from the
powers of the sources. There are $2q$ delta functions from the source
correlators \eq\nr{eq:rhorho} and $2q$ momentum conservation delta
functions from the three gluon vertices. Not all these delta functions
are independent: two of them end up having the same argument, yielding
one factor of the transverse area (denoted $S_\perp$). Therefore,
there is only one remaining transverse momentum integral.  One can
choose this remaining momentum to be the one circulating in all of the
lower part of the diagram, which we shall denote by $\kt$ (this is the
momentum that circulates along the loop in \fig\ref{fig:polymer}).  On
the rainbow side of the diagram there is a squared propagator
$1/(\kt-\pt_i)^4$ for all the sources $i=1 \dots q$. On the
non-rainbow side the transverse momentum in all the propagators is the
same, giving a factor $1/\kt^{4q}$.  Half of these propagators are
cancelled by the squares of the Lipatov vertices, which also
contribute an inverse square of the external momentum. Combining the
combinatorial factors from the source averages, the propagators,
Lipatov vertices and factors from the invariant measure, we finally
get\footnote{ In comparing this to the result 
in~\cite{Dumitru:2008wn}  (\eq(27))  note that there is
an erroneous factor of $1/2$ in \eq(9) of~\cite{Dumitru:2008wn}
which propagates into an additional factor $2^{-2q}$.
In addition, there is an overall error of $(2\pi)^2$. These errors were present
also in Ref.~\cite{Dusling:2009ar}, but have been corrected in the published version.
}
\begin{multline}\label{eq:fullresult}
\left\langle 
\frac{\ud N}{\ud y_1 \ud^2 \pt_1 \dots \ud y_q \ud^2 \pt_q}
\right\rangle_\textrm{conn.}
=
\Big[ 2^q (q-1)! \Big]
\frac{(\nc)^{q}(\nc^2-1) \, S_\perp}{(\pt_1)^2 \cdots (\pt_q)^2}
\frac{
1
}{g^{2q}} \frac{2^q}{(2 \pi)^{3q}}
\\
\times
\int \frac{\ud^2\kt}{(2\pi)^2}
\left(\frac{g^4\mu^{2}(\kt)}{\kt^2}\right)^q
\frac{g^4 \mu^{2}(\pt_1-\kt)}{(\pt_1-\kt )^2} \cdots 
\frac{g^4 \mu^{2}(\pt_q-\kt)}{ (\pt_q-\kt )^2}\;.
\end{multline}
This general formula also reproduces the result of
refs.~\cite{Kovner:1995ts,Kovchegov:1997ke,Gyulassy:1997vt} for the
single inclusive spectrum case $q=1$; in this case the
combinatorial factor in the square bracket must be taken to be 1
instead of 2 to avoid double counting the only contributing diagram.

The weak source result \eq\nr{eq:fullresult} is infrared
divergent in the MV model ($g^4\mu^{2}(\kt)$ constant). Physically 
this is modified by several effects. Even in the weak field limit
BK or BFKL evolution leads to an anomalous dimension $0<\gamma<1$ 
that changes  the behavior into $g^4\mu^2 (\kt) \sim \kt^{2(1-\gamma)}$
in the geometric scaling region $k_\perp \gtrsim \qs$. Deep in the 
saturation regime it has been argued~\cite{Iancu:2002aq}
that the correlator effectively behaves as $g^4\mu^{2}(\kt)\sim \kt^2$.
Ultimately the infrared behavior of the multigluon spectrum 
is regulated by the nonlinear 
interactions that are not included in our present computation. 
This is seen explicitly and analytically in the ``pA'' 
case~\cite{Dumitru:2001ux,Kharzeev:2003wz,Blaizot:2004wu} and in numerical
computations of the glasma fields in the fully nonlinear 
case~\cite{Krasnitz:1998ns,Krasnitz:2000gz,Krasnitz:2001qu,Lappi:2003bi}.
Since the full nonlinear dynamics are known to regulate the
infrared behavior in the case of the single gluon spectrum we have
strong reasons to expect that they will also do so in the case of
multiple gluon production; at the same scale $k_\perp \lesssim \qs$.
We emphasize that an essential point in this argument is that the quantity
appearing in \eq\nr{eq:fullresult} is not a single color charge correlator
divided by a large power $\kt^{2q}$, but the same correlator 
$g^4\mu^{2}(\kt)/\kt^2$ that appears in 
the single inclusive gluon spectrum raised 
to a large power  $q$. 

The effect of saturation  on the multigluon spectrum at $k_\perp \lesssim \qs$
has a very intuitive interpretation in the the 
glasma flux tube picture. The size of the flux
tube, $1/\qs$, is the correlation length of the system and we should
not have contributions from longer distance scales. We effectively
take this into account by regulating all the infrared divergences at
the scale $\qs$, and thus approximating the integral in
\eq\nr{eq:fullresult} by $ \left(2\pi \kappa^{q-1} \qs^{2(q-2)}
  \pt_1^2 \cdots \pt_q^2\right)^{-1}$.  Here $\kappa$ is a constant of
order one that depends on the details of how the infrared divergences
are regulated at the scale $\qs$. Since $\qs$ is the typical momentum of the 
produced gluons, not a lower limit, we expect that numerically $\kappa<1$.
In our analytical calculation we do
not have access to the exact value of this coefficient.
We also use the corresponding approximation for the single inclusive 
spectrum\footnote{We are neglecting an additional logarithmic dependence
in $\pt$.}:
\begin{equation}\label{eq:singleinc}
\left\langle 
\frac{\ud N}{\ud y\ud^2 \pt }
\right\rangle
\approx 
\frac{\nc  (\nc^2-1)}{4 \pi^4 g^2}
\frac{S_\perp (g^2\mu)^{4}}{\pt^4}\;.
\end{equation}
We can now express our result as 
\begin{eqnarray}\label{eq:fullresult2b}
\nonumber
\left\langle 
\frac{\ud N}{\ud y_1 \ud^2 \pt_1 \dots \ud y_q \ud^2 \pt_q}
 \right\rangle_\textrm{conn.} \!\!\!\!\! \!\!\!\!\!
&=&
(q-1)! \,
\frac{(\nc^2-1) \kappa \qs^2 S_\perp}{ 2\pi  }
\left(
\frac{(g^2\mu)^4} {g^2} \frac{1}{2 \pi^3 }
\frac{\nc}{\kappa \qs^2}
\right)^q
\frac{1}{(\pt_1)^4 \cdots (\pt_q)^4}
\\
&=&
(q-1)!  \,
\frac{(\nc^2-1) \kappa \qs^2 S_\perp}{2\pi}
\frac{
\left\langle 
\frac{\ud N}{\ud y_1 \ud^2 \pt_1}
\right\rangle
\dots 
\left\langle 
\frac{\ud N}{\ud y_q \ud^2 \pt_q}
\right\rangle
}{
\left(
(\nc^2-1) \kappa \qs^2 S_\perp  / ( 2\pi )
\right)^q
}
\end{eqnarray}
If we integrate this equation over the rapidities and transverse
momenta of the $q$ gluons, again consistently regulating all the
infrared divergences at the scale $\qs$, we obtain our result for the
factorial cumulant as
\begin{equation}\label{eq:mqresult}
m_q = (q-1)! \,  k \left(\frac{\bar{n}}{k} \right)^q,
\end{equation}
with 
\begin{equation}\label{eq:k}
k =  \kappa \frac{  (\nc^2-1)   \qs^2 S_\perp  }{2\pi}.
\end{equation}
The exact constant factors, encoded in the coefficient $\kappa$,
depend on the exact way the infrared divergences (logarithmic for the
single inclusive, power law for the multigluon correlations) are
regulated. These factors cannot be obtained exactly in an analytic
calculation to the lowest order in the sources.  However, the main
parametric dependences in the relevant variables $\as, \qs, S_\perp$
and $\nc$ can be expected to be the same to all orders in the sources.
A possible additional (mild) $q$-dependence in $\kappa$ would be a
minor correction to the behavior of the probability distribution,
mostly determined by the combinatorial factor $(q-1)!$.

Equations \nr{eq:mqresult} and \nr{eq:k} are the main result of this
paper. One can see that these factorial cumulants \nr{eq:mqresult} are
those that define the \emph{negative binomial} distribution. It arises
very naturally in the Glasma based on the Gaussian combinatorics of
the classical sources and the assumption of the fluctuations in the
system being dominated by a correlation length $1/\qs$.

\section{Glittering Glasma: Interpretation of the result}

A negative binomial distribution is characterized by two parameters,
the mean $\bar{n}$ and $k$, in terms of which the probability to
produce $n$ particles is
\begin{equation}
P_n^{^{\rm NB}}= 
\frac{\Gamma(k+n)}{\Gamma(k)\Gamma(n+1)} 
\frac{\bar{n}^n k^k}{(\bar{n}+k)^{n+k}}\;.
\end{equation}
The distribution is characterized by the generating function
\begin{equation}\label{eq:defgenfunc}
F_{k,\bar{n}}(z) \equiv \sum_{n=0}^{\infty}z^n\,
P_n
= \left( 1 - \frac{\bar{n}}{k} (z-1) \right)^{-k}\;.
\end{equation}
The moments of the distribution can be obtained from the generating
function by differentiating with respect to $z$. The \emph{connected}
parts of the moments, or cumulants, are generated by the logarithm of
the generating function.  The factorial cumulants \nr{eq:defcumul} of
the negative binomial distribution are given by
\begin{equation}
m_q \equiv \left. \frac{\ud^q}{\ud z^q} \ln F_{k,\bar{n}}(z)\right|_{z=1}
= (q-1)! \, k \left(\frac{\bar{n}}{k}\right)^q\; .
\end{equation}
In contrast, for a Poisson distribution $m_1=\bar{n}$ and $m_q=0$ for
$q>1$.  This quantity is the expectation value
\begin{equation}\label{eq:mpexp}
m_q = \langle n(n-1) \cdots (n-q+1)\rangle - \textrm{disc.},
\end{equation}
where the disconnected part ``disc'' can be expressed in terms of the
lower order cumulants. The convention that the expectation value in
\nr{eq:mpexp} is taken of the product $ n(n-1) \cdots (n-p+1)$ and not
of $n^p$ means that we are subtracting the ``Poissonian'' part from
the moment, which is why all the factorial cumulants of a Poisson
distribution are zero for $q\ge 2$. Note that the Poissonian part is
suppressed by powers of $\as$ in the Glasma; thus in practice the
difference between $m_p$ and a conventional cumulant is neglected in
our analysis.

Two common special cases of the negative binomial are the Poisson
distribution, obtained in the limit $k\to \infty$ at fixed $\bar{n}$,
and the geometrical--or Bose-Einstein--distribution obtained when
$k=1$. The negative binomial distribution is wider than a Poisson
distribution typically associated with independent emission of
particles; this can be seen e.g.  from the variance
\begin{equation}
\sigma^2 = \langle n^2 \rangle - \langle n \rangle^2 = 
\bar{n} + \frac{\bar{n}^2}{k}\;. 
\end{equation}

A useful property of the negative binomial distribution with
parameters $\bar{n},k$ is that it is is also the distribution of a sum
of $k$ independent random variables drawn from a Bose-Einstein
distribution with mean $\bar{n}/k$.  This is easily seen from the
generating function in \eq\nr{eq:defgenfunc}, remembering that the
generating function of a sum of independent random variables is the
product of their generating functions\footnote{Explicitly, consider
  $n=n_1 + \dots + n_r$ where the $n_i$'s are independent of each
  other.  The probability distribution of $n$ is then $P_n =
  \sum_{n_1} \cdots \sum_{n_r} \delta(n-\sum_{i=1}^r n_i) P_{n_1}
  \cdots P_{n_r}$ and the generating function $\sum_n z^n P_n =
  \sum_{n_1} \cdots \sum_{n_r} z^{n_1 + \cdots + n_r} P_{n_1} \cdots
  P_{n_r}$, which is the product of the individual generating
  functions for the variables $n_i$.  }.  This has a consequence that
an incoherent superposition of $N$ emitters that have a negative
binomial distribution with parameters $k_0,\bar{n}$ produces a
negative binomial distribution with parameters $Nk_0, N\bar{n}$.  

A natural physical interpretation of our result can be given in terms
of emission from independent glasma flux tubes. Geometrically, the
transverse area $S_\perp$ is filled with $\qs^2 S_\perp$ independent
flux tubes of size $\sim 1/\qs^2$. Each of these tubes emits gluons in
$\nc^2-1$ different colors.  Our result shows that the probability
distribution of gluons of one color emitted from one flux tube is
approximately a Bose Einstein distribution. We do not see an
interpretation of this result as a thermal process, however. It seems
more likely that the distribution is one that maximizes the entropy
(which is the defining property of the BE distribution) because there
is a large number of color sources that emit gluons.  In some sense
the role of a heat bath (large reservoir of energy) in thermodynamics
is played by the large number of color charges resummed into the
effective color current of the CGC.

It is a known experimental
observation~\cite{Arnison:1982rm,Alner:1985zc,Alner:1985rj,Ansorge:1988fg}
(see also \cite{Giovannini:2004yk} for an extensive review)
that multiplicities of charged particles in high energy scattering are
well described as a negative binomial distribution.  Also multiplicity
fluctuations at RHIC have been found to agree with the negative
binomial distribution by the PHENIX
collaboration~\cite{Adler:2007fj,Adare:2008ns}.  Experimentally, the
parameter $k$ increases somewhat with $\delta \eta$, the size of the
rapidity interval in which the particles are measured. The dependence
is, however, very slow for large $\delta \eta$, pointing to the
presence of a long range correlation in the
system~\cite{Giovannini:1985mz}. This is natural in the Glasma
picture, since flux tubes extend over large rapidity intervals. The
number of flux tubes, which gives the parameter $k$ of the negative
binomial distribution, essentially depends only on the transverse area
of the projectiles and on the saturation momentum.

The main difficulty in interpreting the experimental results arises
from the geometrical fluctuations from averaging over different impact
parameters in one finite centrality bin. To minimize this effect one
should use as small centrality bins as possible.  Comparison with a
different method of analysis used by the STAR
collaboration~\cite{Srivastava:2007ei} could be very useful in
disentangling these effects.  To the extent that this uncertainty
allows us to compare results in gold-gold and $p\bar{p}$, the picture
we present seems fairly consistent. For a fixed collision energy we
would expect scaling $k\sim \qs^2 S_\perp \sim \npart$.  While keeping
this caveat in mind, the results from UA5~\cite{Ansorge:1988fg}
and E735~\cite{Lindsey:1991pt}
($k\approx 2 \dots 4$, $\npart = 2$) and PHENIX
$k\approx 350$ for 0-5\% most central ($\npart \approx 350$)
collisions~\cite{Adler:2007fj}
or $k=690$ when extrapolated to a zero centrality 
bin width~\cite{Adare:2008ns}, seem
very consistent with this estimate.  

From the experimental fit of the parameter $k$ in central gold-gold
collisions by PHENIX~\cite{Adare:2008ns} and the value 
$\qs\approx 1.1$~GeV estimated from measurements of the charged multiplicity, 
one can use \eq\nr{eq:k} to 
obtain an estimate $\kappa\approx 0.2$ for the parameter that reflects 
our uncertainty in the infrared sector. 
As we have discussed, it is natural to expect a numerical value of 
slightly less than 1 for $\kappa$.
This, however, means that at RHIC energies the flux tube size, as measured in 
the multiplicity distribution, is not yet very clearly separated
from the confinement scale. At LHC energies we can expect this separation
to be clearer.

For increasing collision energy we would expect $\qs$ and therefore
$k$ to increase. The energies where the UA5 measurements are done are
still in the transition region from a behavior of $k$ decreasing with
energy from lower $\sqrt{s}$, but we would expect $k$ at the LHC to be
clearly larger.  This decreasing behavior at low energy follows
because of the Poisson nature of low energy particle emission, and
that for a Poisson distribution $k \rightarrow \infty$.

The negative binomial has been interpreted as resulting from a partial
stimulated emission or cascade process~\cite{Giovannini:1985mz}. 
It has been known in the literature~\cite{Biyajima:1995yr,Wilk:2006vp} 
that the distribution would naturally arise from a superposition
of subsystems with Bose-Einstein distributions. 
Nevertheless, a popular approach has remained to interpret the 
observations in terms of a
fluctuating number of strings~\cite{Andersson:1983ia}, each producing
particles typically with a Poisson
distribution~\cite{Braun:2000cc,DiasdeDeus:2003ei} (see also
\cite{Dremin:1993sd,Dremin:2000ep,Dremin:2004ts} for a more pQCD based approach).
While the picture of flux tubes in the glasma has many similarities to
ideas in string model phenomenology, the distribution of particles
produced from one flux tube is different. The probability distribution
of gluons from a glasma flux tube is not a narrow Poissonian, but has
very large fluctuations: the glittering of the glasma.

\section{Summary}
The Glasma provides a successful phenomenology of a particle
production in high energy hadronic collisions.  There is now
experimental data on the ridge phenomena that show flux tube
structures in two particle
correlations~\cite{Putschke:2007mi,Daugherity:2008su,Wenger:2008ts}.
In addition, long range correlations of remarkable strength are seen
in heavy ion collisions~\cite{Srivastava:2007ei}.

The Glitter of the flux tube decay may provide a strong tool for
disentangling various descriptions of the flux tubes, since it
naturally leads to a negative binomial distribution for the
multiplicity of produced particles.  However, in order to make a more
convincing case for the origin of the negative binomial distribution
of particle multiplicities, we need a systematic study of bin size
effects on the extraction of the parameter $k$ from various
centralities of heavy ion collisions.
 
 \section*{Acknowledgements}
 The authors gratefully acknowledge conversations with Raju
 Venugopalan.  L. McLerran was supported in part by the Theoretical
 Physics Division at CEA-Saclay, and this work is a product of the
 stimulating intellectual atmosphere there.  The research of L.
 McLerran is supported under DOE Contract No. DE-AC02-98CH10886.  T.
 Lappi is supported by the Academy of Finland, project 126604.  F.
 Gelis is supported in part by Agence Nationale de la Recherche via
 the programme ANR-06-BLAN-0285-01.

\bibliographystyle{h-physrev4mod2}
\bibliography{spires}

\end{document}